\documentclass{article}
\usepackage{epsf}
\def\ga{\gamma}
\def\om{\omega}
\def\cst{\mbox{\it const.}}
\def\la{\lambda}

\begin{document}

\begin{center}
 {\LARGE On geometry behind Birkhoff theorem\\}
 \vskip 3mm
 \it Pavol \v Severa\vskip 2mm
 Dept. of Theoretical Physics, Comenius University,\\
 84215 Bratislava, Slovakia
\end{center}

\section{Area as the affine parameter}

Suppose $N$ is a 2dim spacetime, i.e. a surface with a metric tensor of
signature $(1,1)$. It is fairly easy to find the isotropic geodesics of
$N$ -- we just integrate the isotropic directions. It is perhaps more
interesting to find affine parameters for these geodesics: one can use
the area of a strip between two infinitesimally close integral curves:
 $$\epsfxsize=7cm \epsfbox{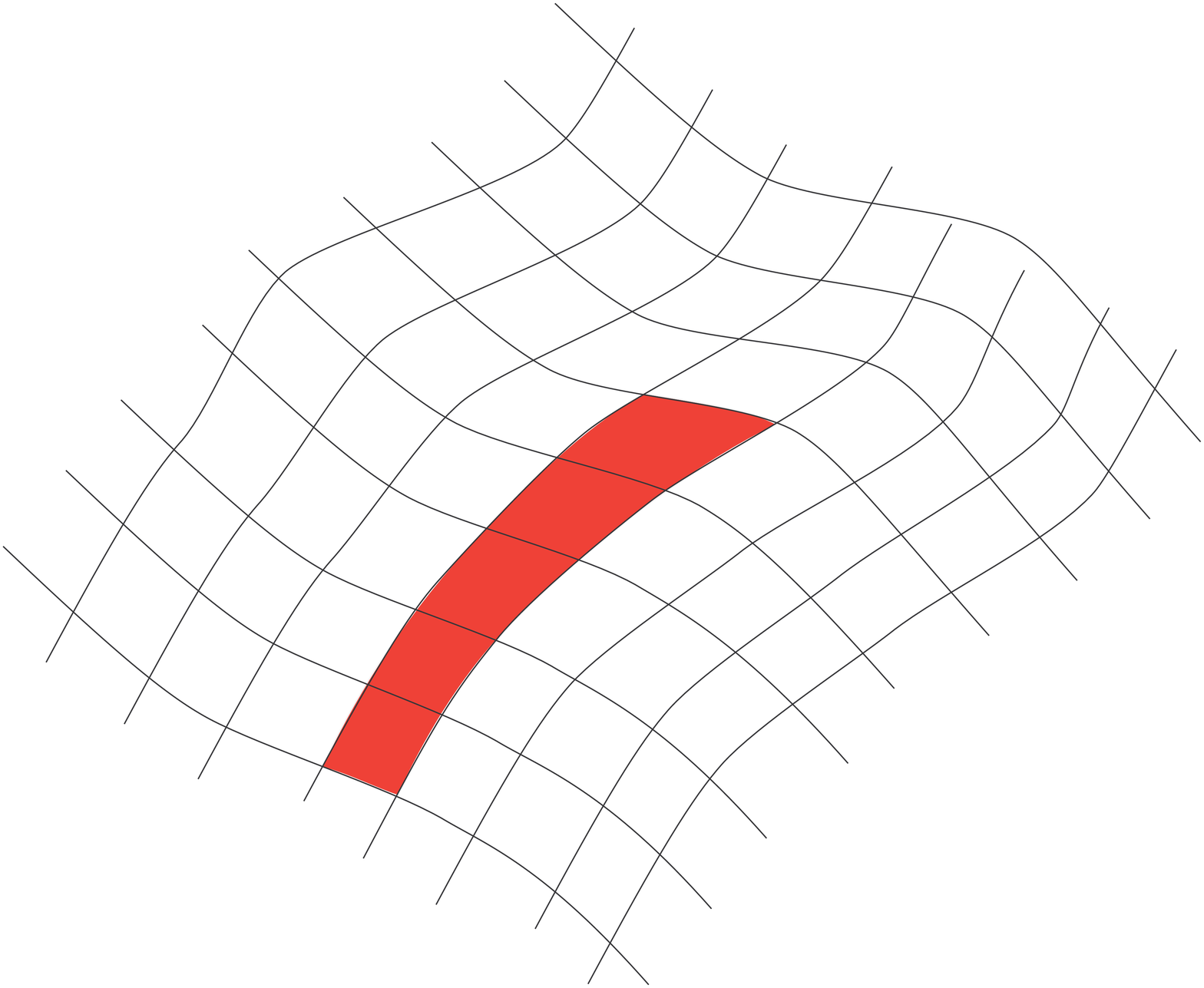}$$

To prove this fact we use the following useful characterization of
geodesics in a pseudo-Riemannian manifold $M$ (for the purpose of
teaching of general relativity, it appears as a convenient  definition):
{\em  If we choose coordinates near a curve $\ga$ so that the metric
tensor is a constant plus $O(r^2)$, where $r$ is (say) the Euclidean
distance to $\ga$, then $\ga$ is a geodesic iff it is a straight line.}
Now our claim that area can be used as the affine parameter is clear,
since it is surely so for constant metric tensor.

We shall use our result in the next section to prove Birkhoff theorem,
but now, as a digression, we mention some other elementary applications.
We can use it conveniently   to
check isotropic geodesic completeness of 2dim spacetimes. Consider, for
example, the Eddington-Finkelstein metrics:
 $$ds^2=-(1-{2m\over r})du^2+2dudr.$$
 We prove that the  geodesic $\ga$ on the following picture is incomplete
to the left; we do not use the exact form of the metrics, only the fact that
it is invariant w.r.t. horizontal translations and that the geodesics converge
on the left (on the picture, $u$ is the horizontal coordinate):

 $$\epsfxsize=7cm \epsfbox{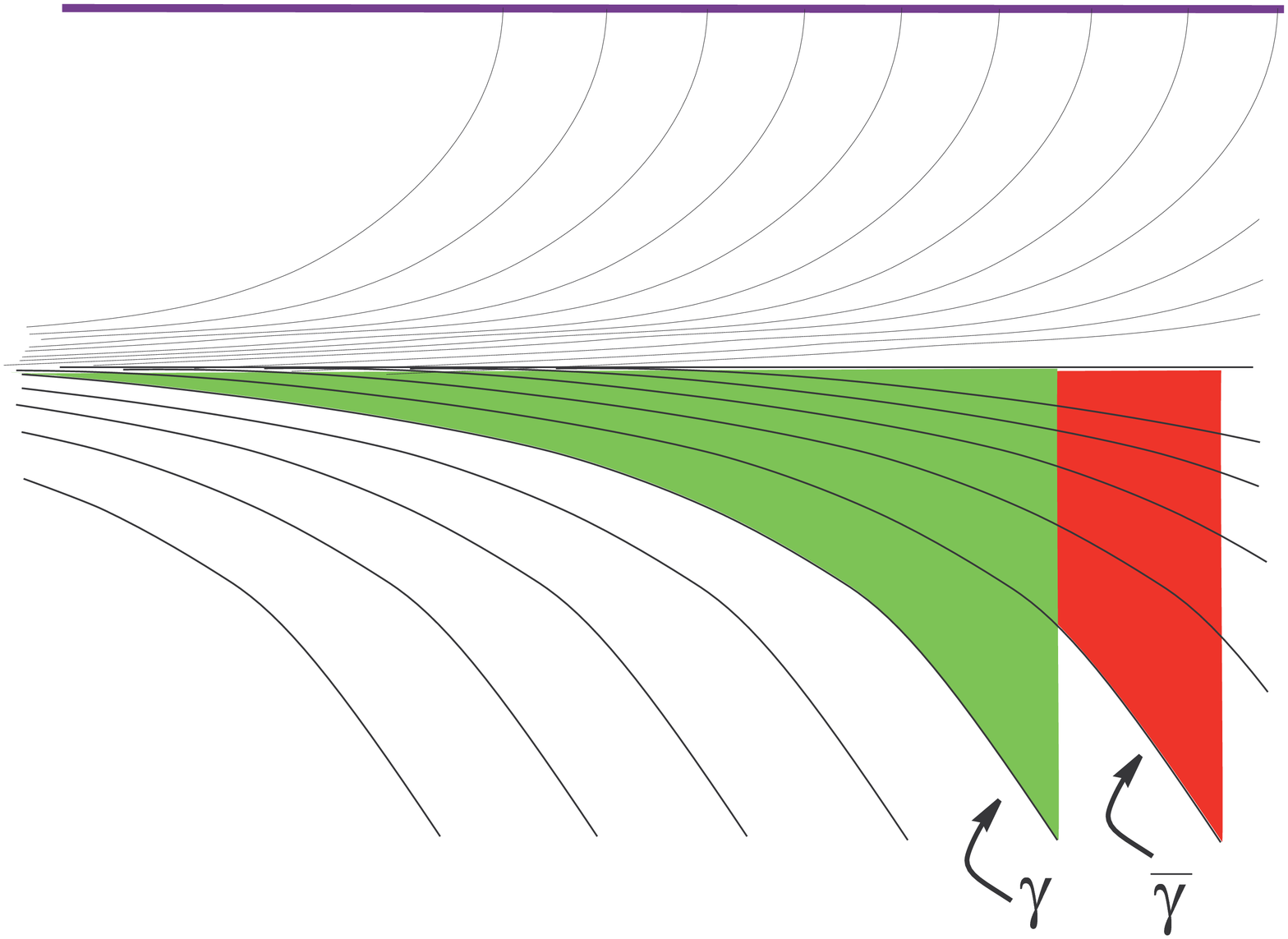}$$  

Indeed, the green area between $\ga$ and $\bar\ga$ is equal to the red area:
to see it,
just take the green triangle and translate it to the right. The red area is
finite, q.e.d.

As another example, consider a $2n$-gon with isotropic sides. Widen each
of its sides to an infinitesimally narrow strip between two isotropic
curves, and compute the expression
 $$\mu={A_1 A_3\dots A_{2n-1}\over A_2 A_4\dots A_{2n}},$$
where $A_i$ is the area at the $i$'th corner, as on the picture:
 $$\epsfxsize=7cm \epsfbox{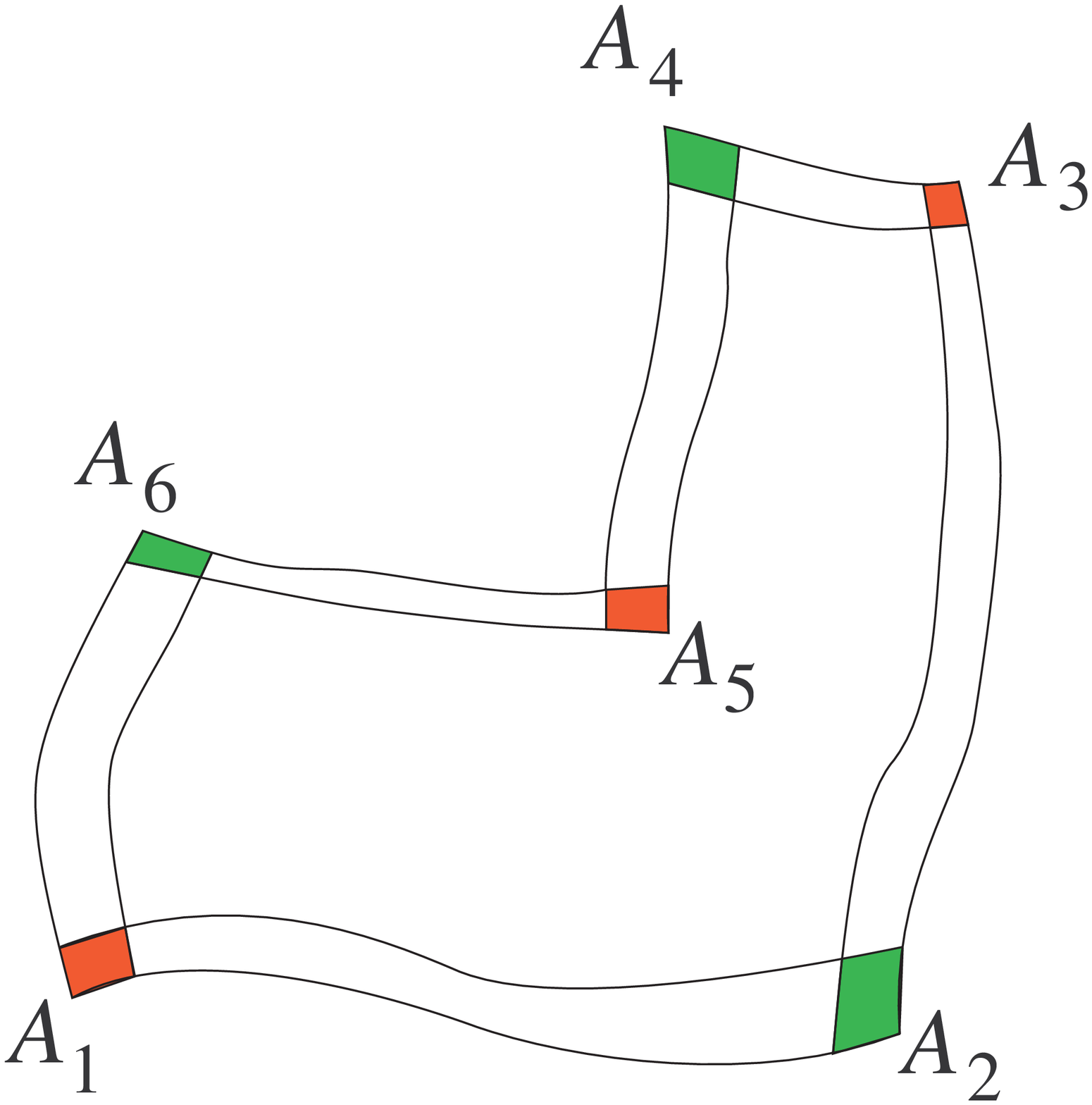}$$
 
$\mu$ is clearly independent of the choice of the widening of the sides:
if we widen the $i$'th side in a different way, $A_i$ and $A_{i+1}$ get
multiplied by the same number, so that $\mu$ doesn't change. We leave it
as an excercise to the reader to prove that $\mu$ is actually the result
of the parallel transport along the polygon, so that it is equal to the
exponential of $\pm$ the integral of  the curvature inside the polygon.

\section{Birkhoff theorem}

\def\ric{\mbox{\it Ric}}
Suppose $M$ is a 4dim spacetime on which $SO(3)$ acts by isometries, so
that all the orbits are spheres. Birkhoff theorem states that under some
assumption on the Ricci tensor, there is a 1-parameter group of isometries
of $M$, commuting with $SO(3)$.

As one easily sees, $M$ is of the form
$M=N\times S^2$, where $N$ is a surface, and
 $$ds^2_M=ds^2_N+r^2ds^2_{S^2},$$
where $r>0$ is a function on $N$ and $ds^2_{M,N,S^2}$ are the metrics on $M$,
 $N$ and the unit sphere. Let $R$ and $\ric$ be the Riemann and Ricci tensor
on $M$, and $\om_N$ the area form on $N$. Finally, let $X_r$ be the Hamiltonian
vector field on $N$ generated by $r$ with symplectic form $\om_N$, i.e.
 $$dr=\om_N(\cdot,X_r).$$ 

{\em Birkhoff theorem: If $\ric(v,v)=0$ for any isotropic $v$ tangent to
$N$, then $X_r$ is a Killing vector field on $M$.}

The proof is split into two lemmas:

{\em Lemma 1: If $N$ is a 2dim spacetime then a vector field $w$ on $N$
is conformal iff for any isotropic geodesic $\ga$, $\om_N(\dot{\ga},w)$
is constant along $\ga$.}

{\em Lemma 2: Under the same assumptions as in Birkhoff theorem, if
$\ga$ is any isotropic geodesic in $N$ and $s$ is its affine parameter,
then $dr/ds=\cst$ along $\ga$.}

{\em Proof of the theorem:} We have to prove that the flow of $X_r$
preserves $r$ and $ds^2_N$. First we show that $X_r$ is conformal on
$N$. It follows from the lemmas, since
 $$\cst=dr/ds=\langle\dot{\ga},dr\rangle=\om_N(\dot{\ga},X_r).$$
 
It remains to check that $X_r$ preserves $\om_N$ and $r$, but it certainly
does, since it is a Hamiltonian vector field q.e.d.

{\em Proof of Lemma 1:}  This follows immediately from our result in
\S1. Indeed, $w$ is conformal iff its flow transports isotropic curves
 to isotropic curves. As we noticed, the area between $\ga$ and an infinitesimally
close curve $\bar{\ga}$ is a linear function of $s$ iff $\bar{\ga}$ is isotropic
q.e.d.

{\em Proof of Lemma 2:} 
If $\ga$ is an isotropic geodesic in $N$ and $P\in S^2$ then $\ga\times
\{P\}$ is a geodesic in $M$. It follows from the $O(2)$-symmetry of 
isometries of $S^2$ preserving $P$: a geodesic is uniquely determined by its
velocity at a point, so that if the velocity is $O(2)$ invariant, the geodesic
must be pointwise $O(2)$-invariant, i.e. it lies in $N\times\{P\}$.

If $Q$ is a different point in $S^2$ then $\ga\times\{Q\}$ is also a
geodesic; therefore, if we take $Q$ infinitesimally close to $P$ we get
that any constant vector $a\in T_PS^2$ satisfies the Jacobi (geodesic
deviation) equation
 $$\ddot{a}+R(a,\dot{\ga})\dot{\ga}=0.$$
On the other hand, the parallel transport along $\ga\times\{P\}$
 must be $O(2)$-equivariant, i.e. it acts as a multiple of the identity
on $T_PS^2$. It also preserves lengths, so that $a/r$ is parallel, and
 $$\ddot{a}=(ra/r)\ddot{\;}=\ddot{r}a/r.$$
 
 To prove that $\ddot{r}=0$ it remains to show that if $v$ is an
isotropic vector tangent to $N$ and $a$ is a vector tangent to $S^2$,
then $R(a,v)v=0$. By definition, $\ric(v,v)$ is the trace of the linear
map $A$ defined by $A(w)=R(w,v)v$. From $O(2)$ symmetry we see that
$A(a)=\la a$ for some number $\la$ (independent of $a$); on the other
hand, $A$ restricted to $TN$ is nilpotent, since $A(v)=0$ and $A(w)$ is
orthogonal to $v$ for any $w$, hence it is a multiple of $v$. Therefore
$0=\mbox{\it Tr}A=2\la$, i.e $A(a)=0$, q.e.d.

\end{document}